\input harvmac
%\draftmode

%%%%%%%%%%%%%%%%%%%%%%%%%%%%%%%%%%%%%%%%%%%%%%%%%%%%%%%%%%%%%%%%%%%%%%%%%
\def\npb#1#2#3{{\it Nucl.\ Phys.} {\bf B#1} (19#2) #3}
\def\npbm#1#2#3{Nucl.\ Phys. {\bf B#1} (20#2) #3}
\def\plb#1#2#3{{\it Phys.\ Lett.} {\bf B#1} (19#2) #3}

\def\atmp#1#2#3{{\it Adv.\ Theor.\ Math.\ Phys.} {\bf #1} (19#2) #3}
\def\jhep#1#2#3{JHEP {\bf #1} (#2) #3}
%%%%%%%%%%%%%%%%%%%%%%%%%%%%%%%%%%%%%%%%%%%%%%%%%%%%%%%%%%%%%%%%%%%%%%%%%

\def\frac#1#2{{#1 \over #2}}
\def\Nfour{{${\cal N} = 4$\ }}
\def\bZ{{\bf Z }}
\def\slz{{$SL(2,\bZ)$\ }}
\def\O{{O_{20'}}}
\def\K{{K_{1}}}
\def\tK{{\tilde K}}
\def\th{{\theta}}
\def\ga{{\gamma}}
\def\x{{\tau_1}}
\def\y{{\tau_2}}
\def\CO{{\cal O}}
\def\Del{{\Delta}}
\def\DO{{\Delta_{\cal O}}}
\def\dl{{\delta}}
\def\dlb{{\bar \delta}}
\def\be{{\beta}}
\def\bed{{\dot \beta}}
\def\bPhi{{\bar \Phi}}
\def\ta{{\tau}}
\def\t0{{\tau_0}}
\def\al{{\alpha}}
\def\ald{{\dot{\alpha}}}
\def\tr{{\rm tr}}
\def\CL{{\cal L}}

\def\CS{{\cal S}}

\def\btau{{\bar \tau}}

\def\bw{{\bar w}}
\def\Ga{{\Gamma}}
%%%%%%%%%%%%%%%%%%%%%%%%%%%%%%%%%%%%%%%%%%%%%%%%%%%%%%%%%%%%%%%%%%%%%%%%%
\lref\bkrsa{M.~Bianchi, S.~Kovacs, G.~C.~Rossi, and Y.~S.~Stanev, ``On the
Logarithmic Behaviour in N=4 SYM Theory,'' \jhep{9908}{1999}{020},
hep-th/9906188.}
\lref\bkrsb{M.~Bianchi, S.~Kovacs, G.~C.~Rossi, and Y.~S.~Stanev,
``Anomalous Dimensions in N=4 SYM Theory at Order $g^4$,''
\npbm{584}{00}{216}, hep-th/0003203.}
\lref\bkrsc{M.~Bianchi, S.~Kovacs, G.~C.~Rossi, and Y.~S.~Stanev,
``Properties of the Konishi Multiplet in N=4 SYM Theory,''
\jhep{0105}{2001}{042}, hep-th/0104016.}
\lref\afp{G.~Arutyunov, S.~Frolov, and A.~C.~Petkou, ``Perturbative and
Instanton Corrections to the OPE of CPOs in N=4 $SYM_4$,''
\npbm{602}{01}{238}; Erratum-ibid. {\bf B609} (2001) 540, 
hep-th/0010137}
\lref\dhkmv{N.~Dorey, T.~J.~Hollowood, V.~V.~Khoze, M.~P.~Mattis, and 
S.~Vandoren, "Multi-Instanton Calculus and the AdS/CFT Correspondence in 
N=4 Superconformal Field Theory," \npb{552}{99}{88}, hep-th/9901128;
N.~Dorey, T.~J.~Hollowood, V.~V.~Khoze, and M.~P.~Mattis, ``The Calculus
of Many Instantons,'' {\it Phys.\ Rept.\ } {\bf 371} (2002) 231,
hep-th/0206063.}
\lref\kovacs{S.~Kovacs, ``N=4 Supersymmetric Yang-Mills Theory and the
AdS/SCFT Correspondence,'' hep-th/9908171.}
\lref\bers{M.~Bianchi, B.~Eden, G.~Rossi, and Y.~S.~Stanev, ``On
Operator Mixing in N=4 SYM,'' hep-th/0205321.}
\lref\appss{G.~Arutyunov, S.~Penati, A.~C.~Petkou, A.~Santambrogio, and
E.~Sokatchev, ``Non-Protected Operators in N=4 SYM and Multiparticle
States of AdS$_5$ SUGRA,'' hep-th/0206020.}
\lref\ryzhov{A.~V.~Ryzhov, ``Quarter BPS Operators in N=4 SYM,''
\jhep{0111}{2001}{046}, hep-th/0109064.} 
\lref\agj{D.~Anselmi, M.~T.~Grisaru, and A.~A.~Johansen, ``A Critical
Behaviour of Anomalous Currents, Electric-Magnetic Universality and
CFT$_4$,'' \npb{491}{97}{221}, hep-th/9601023; D.~Anselmi,
D.~Z.~Freedman, M.~T.~Grisaru, and A.~A.~Johansen, ``Universality of the
Operator Product Expansions of SCFT$_4$,'' \plb{394}{97}{329},
hep-th/9608125; D.~Anselmi, ``The N=4 Quantum Conformal Algebra,''
\npb{541}{99}{369}, hep-th/9809192.}
\lref\maass{H.~Maass, {\it Lectures on Modular Functions of One Complex
Variable,} Tata Institute of Fundamental Research, (1983).}
\lref\terras{A.~Terras, {\it Harmonic Analysis on Symmetric Spaces and
Applications I,} Springer-Verlag, (1985).}
\lref\apostol{T.~M.~Apostol, {\it Modular Functions and Dirichlet Series
in Number Theory,} 2nd Edition, Springer-Verlag, (1990).}
\lref\bump{D.~Bump, {\it Automorphic Forms and Representations,}
Cambridge University Press, (1997).}
\lref\emd{P.~Goddard, J.~Nuyts and D.~I.~Olive, ``Gauge Theories And
Magnetic Charge,'' \npb{125}{77}{1}$\,$; C.~Montonen and D.~I.~Olive,
``Magnetic Monopoles As Gauge Particles?'' \plb{72}{117}{77}$\,$;
H.~Osborn, ``Topological Charges For N=4 Supersymmetric Gauge Theories
And Monopoles Of Spin 1,'' \plb{83}{79}{321}.}
\lref\svw{A.~Sen, ``Dyon-Monopole Bound States, Self-Dual Harmonic Forms
on the Multi-Monopole Moduli Space, and SL(2,Z) Invariance in String
Theory,'' \plb{329}{94}{217}, hep-th/9402032$\,$; C.~Vafa and E.~Witten, ``A
Strong Coupling Test of S-Duality,'' \npb{431}{94}{3}, hep-th/9408074.}
\lref\senb{A.~Sen, ``Descent Relations Among Bosonic D-Branes,'' {\it
Int.\ J.\ Mod.\ Phys.} {\bf A14} (1999) 4061, hep-th/9902105.}
\lref\intril{K.~Intriligator, ``Bonus Symmetries of N=4 Super-Yang-Mills
Correlation Functions via AdS Duality.'' \npb{551}{99}{575},
hep-th/9811047.}
\lref\intski{K.~Intriligator and W.~Skiba, ``Bonus Symmetry and the
Operator Product Expansion of N=4 Super-Yang-Mills,''
\npb{559}{99}{165}, hep-th/9905020.}
\lref\adscft{J.~Maldacena, ``The Large N Limit of Superconformal Field
Theories and Supergravity,'' \atmp{2}{98}{231}, hep-th/9711200;
S.~S.~Gubser, I.~R.~Klebanov, and A.~M.~Polyakov, `` Gauge Theory
Correlators from Non-Critical String Theory,'' \plb{428}{98}{105},
hep-th/9802109; E.~Witten, ``Anti De Sitter Space And Holography,''
\atmp{2}{98}{253}, hep-th/9802150.}
\lref\harvey{J.~A.~Harvey, ``Magnetic Monopoles, Duality, and
Supersymmetry,'' hep-th/9603086.}
\lref\divech{P.~Di Vecchia, ``Duality in Supersymmetric N=2,4 Gauge
Theories,'' hep-th/9803026.}
\lref\arutysok{G.~Arutyunov and E.~Sokatchev, "Implications of
Superconformal Symmetry for Interacting (2,0) Tensor Multiplets," \npbm{635}{02}{3-32}, hep-th/0201145.}
\lref\witten{E.~Witten, "Some Comments on String Dynamics," hep-th/9507121.}
\lref\bgk{M.~Bianchi, M.~B.~Green, and S.~Kovacs, ``Instanton Corrections
to Circular Wilson Loops in N=4 Supersymmetric Yang-Mills,''
\jhep{0204}{2002}{040}, hep-th/0202003.}

%%%%%%%%%%%%%%%%%%%%%%%%%%%%%%%%%%%%%%%%%%%%%%%%%%%%%%%%%%%%%%%%%%%%%%%%%%

\Title{\vbox{\baselineskip12pt
\hbox{hep-th/0212172}
\hbox{EFI-02-49}
\vskip-.5in}}
{\vbox{\centerline{$SL(2,{\rm Z})$\  Multiplets in \Nfour SYM Theory}}}
%\centerline{}}}
\medskip\medskip\bigskip
\centerline{Li-Sheng Tseng}
\bigskip\medskip
\centerline{\it Enrico Fermi Institute and Department of Physics}
\centerline{\it  University of Chicago} 
\centerline{\it  5640 S. Ellis Ave., Chicago, IL 60637, USA}
\centerline{\tt lstseng@theory.uchicago.edu}
\medskip
\baselineskip18pt
\medskip\bigskip\medskip\bigskip\medskip
\baselineskip16pt
\noindent
We discuss the action of \slz on local operators in $D=4$, \Nfour
SYM theory in the superconformal phase.  The modular property of the
operator's scaling dimension determines whether the operator
transforms as a singlet, or covariantly, as part of a finite or
infinite dimensional multiplet under the \slz action.  As an example, we
argue that operators in the Konishi multiplet transform as
part of a $(p,q)$ $PSL(2,\bZ)$ multiplet.  We also comment on the
non-perturbative local operators dual to the Konishi multiplet.

\Date{December, 2002}
%%%%%%%%%%%%%%%%%%%%%%%%%%%%%%%%%%%%%%%%%%%%%%%%%%%%%%%%%%%%%%%%%%%%%%%%

\newsec{Introduction}

The \Nfour super Yang-Mills (SYM) theory in four dimensions is widely
believed to realize an \slz duality \svw.  The duality group acts on
the two parameters of the theory - the coupling, $g$, and the theta angle,
$\th$.  Writing the parameters as $\tau=\x + i \y \equiv
\frac{\th}{2\pi}+i\frac{4\pi}{g^2}$, the \slz action is that of the
modular transformation 
\eqn\slt{\tau \to \tau'=A(\ta)=\frac{a\tau+b}{c\tau+d}~,}
where $A=\left (\matrix{a&b\cr c&d}\right )\in SL(2,\bZ)$.\foot{The
duality group is $PSL(2,\bZ)$, if identifying each matrix with its
negative.}  By duality,
theories with different $\tau$'s that are connected by an \slz 
transformation are physically equivalent.  For $\th=0$
and $a=d=0$, \slt\ reduces to $g\to 4\pi/g$, the weak-strong coupling duality 
of Montonen and Olive \emd.   

Discussions concerning \slz duality in \Nfour SYM theory have mainly focused on
the Coulomb phase of the theory, where the
global $SU(4)\sim SO(6)$ $R$ symmetry is spontaneously broken.  In the
Coulomb phase, duality has provided important insights for understanding
the non-perturbative aspects of the theory.  For example, it implies the
invariance of the BPS mass spectrum under \slt\ (for reviews, see
\refs{\harvey, \divech}).  Such invariance only occurs
if the non-perturbative monopoles and dyonic states are taken into
account.  Indeed, by the BPS mass formulas, the W-bosons, monopoles, and
dyons together are organized into $(p,q)$ \slz multiplets.  Dynamically,
duality also implies that monopoles at strong coupling behave like
W-bosons at weak coupling. 

\Nfour SYM theory has another important phase, the superconformal phase,
where the theory is invariant under the superconformal group $PSU(2,2|4)$, with
$SO(4,2)\times SU(4)$ as the bosonic subgroup.  Here, the observables
consist not of particles and solitons, but locally, operators with definite
scaling dimensions organized into superconformal
multiplets.\foot{Wilson loops, which are non-local observables, will not
be discussed here.  A discussion of Wilson loops and \slz duality can be
found in \bgk.}  In this paper, we explore the action the
\slz duality on the local
observables in the superconformal phase.  Whether an operator is mapped
into itself or to a non-perturbative operator under an \slz transformation is
determined by the invariance of its scaling dimension, as a function of
$\tau$, under the modular transformation.  In general, operators can transform
as an \slz singlet, or as part of a finite or infinite dimensional \slz
multiplet.  

As paradigms, we analyze two superconformal multiplets that have appeared
prominently in the study of $D=4$ quantum conformal algebra \agj\ and
also AdS/CFT correspondence \adscft.  They are the $1/2$-BPS
supercurrent multiplet and the non-BPS Konishi multiplet.
We show that operators in the supercurrent multiplet map into
themselves up to a multiplicative factor similar to that conjectured by
Intriligator \intril .  However, using the perturbative and
non-perturbative calculations for the scaling dimension of the Konishi
operator in \refs{\bkrsa,\bkrsb,\bkrsc}, we argue that the Konishi
multiplet transforms covariantly under the \slz transformation.  In particular,
the Konishi multiplet is the $(1,0)$ element of a $(p,q)$ $PSL(2,\bZ)$
multiplet of non-BPS superconformal multiplets in the \Nfour SYM
theory.\foot{In the AdS/CFT correspondence, the Konishi operator is
expected to be associated with stringy states.  From this perspective,
comments on the covariant transformation of the Konishi operator under
\slz were made in \refs{\bkrsa, \bkrsc}.}  

In section two, we briefly review the superconformal representations of
\Nfour SYM theory and set up our notation.  In section three, we discuss the
implications of \slz duality on the spectrum of operators and examine in
detail the transformation properties of the supercurrent and Konishi
multiplets.  We close in section four with some remarks on modular
functions and non-perturbative duals of the Konishi multiplet.

\newsec{Superconformal representations of the \Nfour SYM theory}

The \Nfour SYM Lagrangian is constructed from the component fields of
the \Nfour gauge multiplet transforming in the adjoint representation of
the gauge group $G$.  For simplicity, we will treat only the case
$G=SU(N)$.  The fields consist of scalars, $\phi^I$, with $I=1,\ldots,
6$ in the {\bf 6} of $SU(4)$ ($R$ symmetry group), complex Weyl spinors,
$\psi_{A\al}$ with $A=1,\ldots,4$ in the ${\bar {\bf 4}}$ of $SU(4)$,
and a gauge field $A_\mu$.  The fields are normalized such that the
action has the form
\eqn\action{\CS=\int\,d^4x\; {\rm Tr}\left\{-\frac{1}{2} F_{\mu\nu}F^{\mu\nu}
-\frac{\th g^2}{16\pi^2}F_{\mu\nu}*F^{\mu\nu} - D^{\mu}\phi^I
 D_{\mu}\phi^I + \ldots \right\}~.}

In the superconformal phase, with $<\!\phi^I\!\!>=0$, the quantum theory is
described by operators that transform under scale transformations with
definite scaling dimensions, $\Del$.  Specifically, the operators are
eigenfunctions of the dilation operator, $D$, with eigenvalue,
$-i\Del$.   Besides its scaling dimension, each operator is also
labelled by its Lorentz and $SU(4)$ representations as required from the
decomposition of the global bosonic symmetry $SO(4,2)\times SU(4)\supset
SO(1,1)\times SO(3,1) \times SU(4)$.

The operators are naturally organized into representations of the
superconformal algebra.  Such a representation module is constructed
starting with a superconformal primary, the lowest weight (scaling dimension)
operator in the module, and then acting on it with the 16 supersymmetry
operators, $Q_\al^A$ and $\bar{Q}_\ald^A$, and momentum operators,
$P_\mu$.  Acting by $Q$ or $\bar{Q}$ increases $\Del$ by 1/2 and
generates conformal primaries while $P_\mu$ increases $\Del$ by 1 and
generates conformal descendants.  The superconformal primary with the
smallest scaling dimension is the identity operator and corresponds to
the trivial one-dimensional representation with scaling dimension $\Del=0$.  In
the free theory with zero coupling, there are two superconformal
primaries with $\Del=2$.  One is the superconformal primary
$O_{20'}=\tr\left(\phi^I\phi^J-{\frac{\delta^{IJ}}{6}\sum_{L=1}^{6}\phi^L\phi^L}\right
)$, the lowest weight operator of the supercurrent multiplet.  This
multiplet is a 1/2
BPS short multiplet with the representation generated by only 8
supersymmetry generators.  Being BPS, the scaling dimension of the
supercurrent multiplet is ``protected'' or remains unchanged for all
values of $\tau$.  Operators in this multiplet include the
$SU(4)$ $R$-current, $J^{IJ}_{\al\ald}= \dl\dlb\O$, the supercurrents,
$S^A_{\al\be\ald}=\dl^2\dlb\O$ and ${\bar S}_{A\al\ald\bed}=\dl\dlb^2\O$, and
the energy-momentum tensor, $T_{\al\ald\be\bed}= \dl^2\dlb^2\O$.  Here, we
follow the notation in \intril\ where, for example,
$\dl\dlb\O=\{Q,[\bar{Q},\O]\}$.  The other dimension two superconformal
primary is the Konishi operator, $\K = \sum_{K=1}^{6}\tr\;\phi^K\phi^K$,
the lowest weight operator of the non-BPS Konishi multiplet.  The free
theory chiral
current is a member of this multiplet.  The scaling dimension of this long
multiplet is not protected and the operator has a nonzero anomalous
dimension when the theory is interacting.  
     
\newsec{\slz invariance of the superconformal \Nfour SYM theory}

In the Coulomb phase of the theory, \slz duality implies the invariance
of the BPS mass spectrum under modular transformation of $\tau$.  However,
in the superconformal phase, operators are labelled by their scaling
dimensions and their Lorentz and $SU(4)$ representations.  Since the values
of the Casimirs of Lorentz and $SU(4)$ representations are discrete
and not continuous, an operator's Lorentz and $SU(4)$
representations can not vary with $\tau$ or under \slz transformation.
As for scaling dimensions, we can consider the spectrum of scaling
dimensions for all operators in the theory for each value of $\tau$.  Duality
then implies that the scaling dimension spectrum is invariant under
the transformation of \slt.

The invariance of the scaling dimension spectrum constrains the
transformation properties of operators under $SL(2,\bZ)$.
Consider the theory at a specific value of $\tau$.  For a conformal
primary operator
$\CO_{\tau}$ with scaling dimension $\DO(\x,\y)$, the two-point
correlation function is fully determined by conformal invariance to be 
\eqn\twocor{<\CO_{\tau}(x_1)\CO_{\tau}(x_2)>_{\tau}\;\sim
\frac{1}{|x_1-x_2|^{2\DO(\x,\y)}}} 
where we have ignored any constant factor that can be absorbed in the 
normalization of $\CO_{\tau}$.  In general,
both $\CO_{\tau}$ and $\DO(\x,\y)$ may have non-holomorphic dependence
on $\tau$.  Now under an \slz transformation with $\tau\to\tau'$,
duality implies the existence of a primary operator
$\CO'_{\tau'}$ in the theory at $\tau'$ that has the scaling dimension
$\DO_{'}(\x',\y')=\DO(\x,\y)$.  Explicitly,
\eqn\twocord{<\CO'_{\tau'}(x_1)\CO'_{\tau'}(x_2)>_{\tau'}\;\sim \;
<\CO_{\tau}(x_1)\CO_{\tau}(x_2)>_{\tau}\;\sim
\frac{1}{|x_1-x_2|^{2\DO(\x,\y)}}~.}
where $\CO'_{\tau'}$ by \slz invariance must have the same Lorentz and $SU(4)$
representations as $\CO_{\tau}$.  Note that \twocord\ must hold true
for any values of $\tau$ and $\tau'$ related by an \slz
transformation.

Now if the scaling dimension satisfies the modular invariance
condition $\DO(\x',\y')=\DO(\x,\y)$, then we simply have
$\CO'_{\tau'}\sim\CO_{\tau}$.\foot{We assume that there is no
degeneracy of operators having identical global symmetry representations
and scaling dimensions for all $\tau$.  Degeneracies of non-BPS
operators that arise at $g=0$ are typically broken by operator mixing at
nonzero coupling.}  Therefore, if the
operator's scaling dimension is a modular function, (i.e.~a function
invariant under the modular transformation of \slt\foot{In the
mathematical literature, the term modular
function sometimes refers only to a meromorphic function of $\tau$ that are
invariant under the modular group.  Here, we call any holomorphic or
non-holomorphic function $f(\x,\y)$ modular invariant if simply
$f(\x,\y)=f(\x',\y')$.}~), \slz transforms the
operator into itself, up to a possible multiplicative factor.  This is
the case for all BPS operators which have constant scaling dimensions.
However, if the scaling dimension is not a modular function, then \slz
transformation will act non-trivially on the operator.  The operator
must necessarily transform covariantly as part of a multiplet under
$SL(2,\bZ)$.

Although \slz is an infinite dimensional discrete group, the \slz
multiplet, in general, need not be infinite dimensional.  It is possible
that the scaling dimension is invariant under a subgroup, $\Ga\subset
SL(2,\bZ)$.  If $\Ga$ has finite index in $SL(2,\bZ)$, then the \slz
multiplet will be finite dimensional.  In fact, \slz has infinitely many
finite index subgroups (see \refs{\maass, \terras} and references
therein).  Well-known examples are the
principal congruence subgroup of level $N$, $\Ga(N)$, defined by
\eqn\subgp{\Ga(N)=\left\{ \left(\matrix{a&b\cr c&d}\right) \in SL(2,\bZ) 
\Big{\arrowvert} a\equiv d\equiv 1~{\rm mod~}N,\; b\equiv c\equiv 0~{\rm mod~}N \right \} }
with index
\eqn\subindex{ \left[ SL(2,\bZ):\Ga(N) \right] = N^3 \prod_{n|N} (1 -
n^{-2}) }
where the product is over positive integers $n>1$ that divide $N$.
Nevertheless, if the index is not finite or if the scaling dimension is
not invariant under any element of $SL(2,\bZ)$, then the multiplet will be
infinite dimensional. 

It is worthwhile to point out a simple toy model exhibiting similar
characteristics of conformal operators transforming under duality.  This
is the two dimensional Gaussian model ($c=1$ closed bosonic string
theory) on a circle with Lagrangian density $\CL\sim\partial X
\bar{\partial} X$.  Here, the discrete duality group is the
$\bZ_2$ of T-duality, inverting the radius $R\to 1/R$.  Operators with
conformal dimension invariant under the $\bZ_2$ action map to themselves
up to a negative sign under T-duality.  For example, $\CL\to -\CL$
because $X_R\to -X_R$ under T-duality.  Operators with conformal
dimensions not invariant under radial inversion are transformed into other
states.  This results in the duality mapping between momentum and winding
modes.

Below, we analyze the \slz transformation property of the supercurrent
and Konishi multiplets in detail to gain more insights on the action of
\slz on superconformal multiplets.  Note that the scaling dimensions of
all operators in a multiplet are determined by the scaling dimension of
the superconformal primary.  Therefore, the study of the scaling
dimension of the primary will determine the \slz multiplet structure
for all operators in the superconformal multiplet.

\subsec{\slz action on the supercurrent multiplet}

Since $\Del_\O$ is a constant, operators in the supercurrent multiplet
map into themselves up to a multiplicative factor under \slz
transformation.  The multiplicative factor differs for different
elements of the multiplet and in general can depend on $\tau$.  For some
of the operators in the multiplet, the transformation factors have
physical significance and can be simply deduced.

Consider first the $R$-current, $J^{IJ}_{\al\ald}=\dl\dlb\O$,
and the energy momentum tensor, $T_{\al\ald\be\bed}= \dl^2\dlb^2\O$.
They are associated with the $SO(1,1)\times SO(3,1) \times SU(4)$
symmetry charges of the theory.  Since these charges are invariant under
\slz transformation, the multiplicative factor for both must be trivial
and they transform invariantly.

Also of importance is the dimension four operator $\Phi=\dl^4\O$ and its
complex conjugate $\bPhi=\dlb^4\O$.  They are the exactly
marginal operators invariant under all 16 supersymmetry generators and
identified with the on-shell Lagrangian density, $\CL\sim
Im[\frac{\tau}{\y}\Phi]$.\foot{Explicitly, $\Phi\sim {\rm Tr}(F^2 + i
*FF)$, from applying on-shell supersymmetry transformation relations.}
As in the simple toy Gaussian model on a circle, where $\CL$ is negative of
itself under T-duality, the \Nfour SYM Lagrangian density also picks up
a non-trivial factor under \slz duality.  This factor can be
obtained as follows. 

As marginal perturbation, $\Phi+\bPhi$ changes the coupling
$g$ of the theory while $\frac{1}{i}(\Phi-\bPhi)$ changes $\th$.  Let us
consider a theory with parameter $\tau$ perturbed by 
\eqn\Lpert{\eqalign{\dl\CL=&\frac{\dl\y}{\y}(\Phi+\bPhi)+\frac{\dl\x}{i\y}(\Phi-\bPhi)\cr
 =&\frac{1}{i\y}[\dl\tau\Phi - \dl\btau\bPhi]~,}}
where $\dl\tau=\dl\x+i\dl\y$ and its complex conjugates are constants
parameterizing the perturbation.  Under the marginal perturbation,
$\tau\to\tau+\dl\tau$.  Now apply the \slz duality to the theory with
the perturbation included.  The dual theory at
$\tau'+\dl\tau'$ is the theory at $\tau'$ perturbed by a dual perturbation
\eqn\Lppert{\dl\CL'=\frac{1}{i\y'}[\dl\tau'\Phi' - \dl\btau'\bPhi']~,}
where $\dl\tau'=\frac{\dl\tau}{(c\tau+d)^2}$.  But since $\dl\tau$ is a
constant and does not transform under $SL(2,\bZ)$, $\Phi$ and $\bPhi$
must pick up a factor under \slz transformation.  From duality, the
transformation is required to be
\eqn\otrans{\Phi\to\frac{1}{(c\tau+d)^2}\Phi \qquad{\rm and}\qquad
\bPhi\to\frac{1}{(c\btau+d)^2}\bPhi~.}
Thus, $\Phi$ and $\bPhi$ transforms with modular weight $(-2,0)$ and
$(0,-2)$, respectively, under modular transformation.\foot{An operator
$O(\tau,\btau)$ with modular weight $(w,\bw)$ transforms under the
modular transformation as $O(\tau,\btau) \to (c\tau+d)^w
(c\btau+d)^{\bw} O(\tau,\btau)$}

Though the above conformal primary operators are in the same
superconformal multiplet, they transform differently under duality.
This implies that the action of \slz and that of the supersymmetry
generators, $\dl$, $\dlb$ do not commute.  Intriligator, in \intril, has
conjectured that all BPS operators transform under \slz duality with a
particular modular weight given by the $U(1)_Y$ charge of the operator.
The $U(1)_Y$ is an outer automorphism of the \Nfour superconformal
algebra that only acts on the fermionic generators.  The conjecture is
motivated by the AdS/CFT correspondence where the $U(1)_Y$ is identified
with the compact $U(1)$ of the $SL(2,{\bf R})$ symmetry in the type IIB
supergravity action.  However, $U(1)_Y$ is broken for non-zero
coupling and its applicability for \slz duality still needs to be
clarified.

\subsec{\slz action on the Konishi multiplet}

Being a long multiplet at non-zero coupling, the scaling dimension of
the Konishi multiplet is not constant with respect to $\ta$.  Explicit
calculations have been carried out to determine both the perturbative and 
non-perturbative contributions to the anomalous dimension of the Konishi
operator, $\ga_\K = \Del_\K -2$, for non-zero $g$ and $\th$.
From perturbative calculations in \refs{\agj, \bkrsa, \bkrsb}, it is
known up to order $g^4$ that
\eqn\anomK{\eqalign{\ga_\K (\tau) =\; & \frac{3N}{4\pi^2}g^2 -
\frac{3N^2}{16\pi^4}g^4 + \ldots \cr =\; & \frac{3N}{\pi}\left
(\frac{1}{\y}-\frac{N}{\pi}\frac{1}{\y^2}+\ldots \right )~,}}
where again $\tau=\x + i \y\equiv\frac{\th}{2\pi}+i\frac{4\pi}{g^2}$.  
As for the dependence on $\th$, note that $\th$ only appears in the
Lagrangian coupled to the surface term $*FF$.  For correlation
functions, $\th$ dependence is known only to arise from instanton
sectors.  Moreover, it was found in \refs{\bkrsa, \afp, \bkrsc} that
non-perturbative instanton effects do not contribute to $\ga_\K$.  This
is technically due to the inability of the two-point function of $\K$ to
provide the necessary
fermion zero modes to match those of the instanton background (see \bkrsc\ and
also \dhkmv\ for details).  Thus, assuming only instanton effects may
give a $\th$ dependence to the scaling dimension, we conclude that
$\ga_\K$ is independent of $\th$.\foot{We assume that no
other non-perturbative effect contributes to the $\th$ dependence of
$\Del_\K$.}

One can ask whether $\Del_\K = 2 + \ga_\K$ with no $\x$ dependence can
possibly be a modular function.  Indeed, one can prove that any modular
function with no $\x$ dependence must be a constant. 
\bigskip
\noindent{\it Theorem}$\,$: Let $f(\ta)$ with $\ta=\x + i\y$ be a
function on the upper half plane, i.e. $\y>0$.  If $f(\ta)$ is a modular
invariant function and is also independent of $\x$ , then $f(\ta)$ is a
constant function.
\medskip
\noindent{\it Proof}$\,$:  With no dependence on $\x$, $f$ is a function
of only one variable $f(\y)$.  Now modular transformation of
$\ta\to\ta'=\frac{a\ta+b}{c\ta+d}\,$ implies $\y \to 
\y'=\frac{\y}{|c\ta+d|^2} $.  Therefore, $f(\y)$ being a
modular invariant function must satisfy
\eqn\fmod{f(\y)=f\left (\frac{\y}{(c\x+d)^2+c^2\y^2}\right ) } 
for any $A=\left (\matrix{a&b\cr c&d}\right )\in SL(2,\bZ)$ and for any
$\x$ on the RHS of \fmod.  We will show that for any $A$ with $c\neq0$,
\fmod\ requires $f(\y)$ is a constant.

First, choose $\x=-\frac{d+x}{c}$ and $\y=\frac{1}{|c|}$, \fmod\ becomes 
\eqn\fmodd{f\left(\y=\frac{1}{|c|}\right)=f\left (\y=\frac{1}{|c|(x^2
+1)}\right )} for any real $x$.  For $0\leq x < \infty $, \fmodd\ implies
$f(\y)=f(\y=1/|c|)$ for all $\y < 1/|c|$.  Now, setting $\x=-d/c$ and
$\y=\frac{1}{|cx|}$ in \fmod , we obtain
$f(\y=\frac{1}{|cx|})=f(\y=\frac{|x|}{|c|})$.  Taking $1\leq x <
\infty$, we conclude that $f(\y)=f(\y=1/|c|)$ for all $\y$. 
\bigskip 

By the above theorem, $\Del_\K(\y)$ can not be a modular function.  This
implies that $\K$ does not transform as a singlet under the \slz duality 
action.  For example, from the $S$ transformation, $\tau\to -1/\tau$, there
must exist a non-perturbative operator, $\K'$ that has scaling dimension
$\Del_{\K'}=2$ as $g^2 \to \infty$.  And because $\Del_K$ is not
invariant under $S$, $\K'$ can not be proportional to
$\K$.  More generally, from the proof of the above theorem, we know that
$\Del_\K(\y)$ is not invariant under any element $\left (\matrix{a&b\cr
c&d}\right )\in SL(2,\bZ)$ with $c\neq0$.  Hence, $\K$ must be an element
in an infinite dimensional multiplet of $SL(2,\bZ)$ which we will call $\tK$.
Since the \slz transformation of $\y$ depends only on the two relatively
prime integers $(c,d)\,$,
elements in $\tK$ can be labelled by a pair of integers, $(p,q)$, with
$p$ and $q$ relatively prime.  The (1,0) and (0,1) elements are
respectively $\K$ and $\K'$.  This representation is similar to that of
the BPS $(p,q)$-string in Type II string theory.  However, for non-BPS
\slz multiplets, the
values of $p$ and $q$ do not correspond to any quantized $U(1)$ charges.
That is in $\tK$, both $(1,0)$ and $(-1,0)$ elements should be
identified with the Konishi operator.  Thus, the $(p,q)$ representation
is more accurately that of $PSL(2,\bZ)$.  This allows the imposition of
the constraint that $p$ be strictly non-negative. 

We can easily write down the scaling dimensions of elements in $\tK$ in
the small $g^2$ expansion.  For each $(p,q)$ element in $\tK$, the
scaling dimensions is given by
\eqn\sdtK{\Del_{(p,q)}(\ta)=2+\frac{3N}{\pi}\left
[\frac{|p+q\ta|^2}{\y}-\frac{N}{\pi}\left (\frac{|p+q\ta|^2}{\y}\right 
)^2+\ldots \right ]~,}
where we have simply applied a modular transformation to \anomK\ by 
replacing $\y$ with $\frac{\y}{|p+q\ta|^2}$.  We expect that the scaling 
dimensions of all elements in $\tK$ with the exception of $\K$ goes to 
infinity as $g^2\to 0$ ($\y\to\infty$).  Thus, in the small coupling 
regime, $\K'$ and other elements of $\tK$ are highly non-perturbative.

The above statements for $\K$ also applies for all other operators in the
Konishi multiplet.  The scaling dimensions of the $(p,q)$ element for
an operator in the multiplet is that of \sdtK\ after replacing the
Konishi operator's canonical dimension with that of the operator of
interest.

\newsec{Discussion}

We have demonstrated that local operators in \Nfour SYM
theory in the superconformal phase may transform non-trivially under
\slz duality.  How an operator transform is
determined by the modular property of its scaling dimension function.
If the function is modular invariant, then it is a singlet under the
transformation.  Otherwise, it sits in a finite or infinite dimensional
multiplet of $SL(2,\bZ)$.  A class of singlets under \slz are operators that
have constant scaling dimensions.  It would be interesting to
identify perturbative operators that have non-trivial modular functions
for their scaling dimensions.  In the theory of automorphic forms, modular
functions that are eigenfunctions of the Laplacian on the
upper half plane have been classified.\foot{The non-Euclidean Laplacian is
$L=\y^2\left(
\frac{\partial^2}{\partial\x^2}+\frac{\partial^2}{\partial\y^2}\right
)$.  The eigenfunctions are known to be of three types: constant,
holomorphic, and non-holomorphic.  We point out that although the
eigenfunctions are by construction modular invariant, modular invariant
functions are generally not eigenfunctions of the Laplacian.  For
references on modular functions, see \refs{\terras, \apostol, \bump}.}     
Taking into account of unitarity constraints which set a lower bound on the
scaling dimension, a class of candidate modular scaling
dimension functions is the non-holomorphic cusp forms.\foot{In
particular, the holomorphic modular functions are not bounded from below
and the non-holomorphic Eisenstein series are not finite as $g \to 0$.}
Although no explicit form of these functions exists, they do exhibit
characteristics of $\x$ dependence similar to those arising from instanton
effects.

Without instanton contributions, a non-constant scaling dimension
can not be modular invariant.  Even though the scaling dimension is invariant
under $T$ transformation, $\tau\to\tau+1$, the lack of $\x$ dependence
requires that the operator in question transform as an $(1,0)$ element
in an infinite dimensional $(p,q)$ $PSL(2,Z)$ multiplet.  This is the
case for the operators in the Konishi multiplet.  As a
corollary, any operator that transforms in a finite dimensional \slz
multiplet must have a non-trivial $\x$ dependence.  At present, no
operator is known to transform in a finite \slz multiplet.  
Nevertheless, it certainly would be interesting for such operators to appear 
or to prove that they are forbidden in the \Nfour SYM theory.

As for the $(p,q)$ multiplet, it consists almost exclusively of
non-perturbative local operators whose fundamental roles arise at the
large coupling regimes.  (The exception is the $(1,0)$ element.)  This
is evident from taking the \slz dual of
perturbative operator product expansions (OPEs) involving the Konishi
operator.  Consider the OPE of two $\O$'s.  Schematically, it is given
perturbatively by
\eqn\ope{\O(x_1)\O(x_2)\to \frac{c}{(x_{12})^4} + \frac{\O}{(x_{12})^2}
+ \frac{\K}{(x_{12})^{2-\ga_\K(\y)}} +\cdots}   
where $c$ is proportional to the central charge, and we have ignored all
$SU(4)$ indices and other proportionality constants.  Under \slz duality,
$\O$ is invariant while $\K$ transforms into an element in $\tK$.  Thus,
for example, at the large $g$ coupling limit with $\th=0$, the OPE's of
two $\O$'s contains the $(0,1)$ operator, $\K'$.  We point out that
the structure constant of two 1/2 BPS short operators and a long
operator, $c_{SSL}$, in general depend on $\tau$.  Thus, even with the
aid of \slz duality, understanding the interactions of $\K'$ at
perturbative coupling will require some knowledge of the dynamics of
$\K$ at strong coupling.

Obtaining a physical understanding of the
non-perturbative $(p,q)$ operators at finite small coupling is
challenging.  Because these operators are non-BPS, the $(p,q)$ labels
are just labels and do not pertain to any symmetry charges.  It may be
possible that a better
understanding may be obtained from a more geometric perspective of \slz
duality, as in the toroidal compactification of the $D=6$, ${\cal
N}=(2,0)$ superconformal theory down to the superconformal \Nfour SYM
theory \witten.  Unlike the \Nfour theory, the corresponding
Konishi-like operator in the ${\cal N}=(2,0)$ theory is found in a
discrete series unitary representation of the superconformal algebra
\arutysok.  One may hope that the subtleties of the toroidal
compactification will reveal the origin of the $(p,q)$ operators and
provide other insights into \slz duality in the superconformal phase.
We leave these questions for future investigations.

\bigskip\medskip\noindent 
{\bf Acknowledgements:}
I am grateful to J. Harvey for helpful discussions and
valuable comments on the manuscript.  I would like to thank R. Kottwitz for
discussions on automorphic forms and S. Sethi for discussions on
non-perturbative aspects of the \Nfour theory.  I have also
benefitted from discussions with R. Bao, B. Carneiro da Cunha,
A. Gopinathan, P. Kraus, E. Martinec, S. Minwalla, K. Okuyama, and
A. Recknagel.  This work was supported in part by NSF
grant PHY-0204608.

\listrefs
\end